# Spin-Coupling Topology in the Copper Hexamer Compounds $A_2Cu_3O(SO_4)_3$ (A=Na, K)

A. Furrer[1], A. Podlesnyak[2], J. M. Clemente-Juan[3], E. Pomjakushina[4], and H. U. Güdel[5]

[1] Laboratory for Neutron Scattering, Paul Scherrer Institut, CH-5232 Villigen PSI, Switzerland

[2] Neutron Scattering Division, Oak Ridge National Laboratory, Oak Ridge, Tennessee 37831, USA

[3] Instituto de Ciencia Molecular (ICMol), Universitat de Valencia, c/Catedrático José Beltrán, 2, 46980 Paterna, Spain

[4] Laboratory for Multiscale Materials Experiments, Paul Scherrer Institut, CH-5232 Villigen PSI, Switzerland

[5] Department of Chemistry and Biochemistry, University of Bern, CH-3000 Bern 9, Switzerland

*Abstract:*

The compounds $A_2Cu_3O(SO_4)_3$ (A=Na, K) are characterized by copper hexamers which are weakly coupled to realize antiferromagnetic order below $T_N \approx 3$ K. They constitute novel quantum spin systems with S=1 triplet ground-states. We investigated the energy-level splittings of the copper hexamers by inelastic neutron scattering experiments covering the entire range of the magnetic excitation spectra. The observed transitions are governed by very unusual selection rules which we ascribe to the underlying spin-coupling topology. This is rationalized by model calculations which allow an unambiguous interpretation of the magnetic excitations concerning both the peak assignments and the nature of the spin-coupling parameters.

# I. INTRODUCTION

Quantum spin systems have been attracting much attention due to numerous features which cannot be interpreted by conventional spin models [1,2]. The search for novel magnetic materials has often been inspired by observations on naturally occuring minerals [2,3]. Of particular interest are compounds which fulfill some of the following conditions: (i) The spin quantum number S of the magnetic ions is low, *i.e.*, S=1/2 or 1; (ii) the dimensionality d of the magnetic system is low, *i.e.*, d=1 or 2; (iii) the connectivity of the network of magnetic ions (*i.e.,* the number of spins to which each spin is coupled) is low; (iv) the interactions between the magnetic ions are geometrically frustrated. Here we focus on the minerals with chemical formula $A_2Cu_3O(SO_4)_3$ found in sublimates of the Tolbachik fission eruption in the years 1975-1976 (A=K) [4] and 2014-2015 (A=Na) [5], which fulfill the criteria (i)-(iii). They are built up of edge-shared tetrahedral spin clusters consisting of six $Cu^{2+}$ ions with S=1/2 spins as schematically shown in Fig. 1. The copper hexamers are weakly coupled giving rise to antiferromagnetic order below $T_N$=3.1 K for A=K [6]. The fascinating properties of the compounds $A_2Cu_3O(SO_4)_3$ have very recently been recognized by Fujihala *et al.* [7], who carried out experimental studies for A=K by magnetic susceptibility, magnetization, heat capacity, and inelastic neutron scattering (INS) measurements (restricted to energy transfers below 2.5 meV). Based on an analysis of the thermodynamic magnetic properties, the presence of a spin triplet ground-state was suggested, which would put the compound into a novel cluster-based Haldane state. The S=1 ground state was later confirmed by Furrer *et al.* [8] by INS experiments performed for the compounds $A_2Cu_3O(SO_4)_3$ (A=Na, K) with energy transfers up to 45 meV.

The spin excitation spectrum of the copper hexamers is an essential ingredient to understand the magnetic properties of the title compounds. However, the existing information reported in Refs. [7,8] is incomplete.

Therefore, we extended the experimental range of the INS spectra up to 90 meV. Phonon and magnetic excitations coexist at these energies, leading to a superposition in the INS spectra. We develop and present a technique, which allows a separation of the two contributions, thus revealing hitherto unobserved magnetic excitations which turn out to be essential in the analysis and interpretation of the data.

INS intensities play a central part in the present study. They have so far not been considered in the previous INS studies of the title compounds [7,8]. Here we show that the very unusual distribution of INS intensity on the various transitions can be ascribed to the spin-coupling topology, which leads to a deeper understanding of the nature of the exchange coupling in the copper hexamers.

## II. EXPERIMENTAL

### A. Sample Preparation and Characterization

Polycrystalline samples of $A_2Cu_3O(SO_4)_3$ (A=Na, K) were synthesized by a solid-state reaction process as described in Ref. [8]. The compounds crystallize in the monoclinic space group C2/c. The samples were characterized by X-ray and neutron diffraction, confirming their single-phase character. The results of the structure refinements were summarized in Ref. [8]. Table I lists the Cu-O-Cu bond angles and Cu-Cu bond distances associated with the copper hexamers.

The $Cu^{2+}$ ions within the hexamers are strongly coupled by magnetic superexchange provided by two oxygen ions located at the centers of the Cu tetrahedra. The Cu hexamers are coupled by weak superexchange through $SO_4$ brigdes along the y- and z-directions, building distinct layers parallel to the yz-plane. These layers are separated by A-planes (A=Na, K), so that the coupling along the x-direction is expected to be negligibly small. The two-dimensional nature of the title compounds was already proposed in Refs. [4,5].

Since the phonon scattering is expected to be dominant in the INS experiments, we tried to synthesize non-magnetic isostructural compounds as phonon standards. The $Cu^{2+}$ ions in the title compounds were replaced by $Zn^{2+}$ ions, however, X-ray diffraction revealed a complete absence of the anticipated $A_2Zn_3O(SO_4)_3$ phase. In fact, there is no report on a successful synthesis of $A_2Zn_3O(SO_4)_3$ in the literature. Therefore, an alternative method based on intensity ratios was developed in order to separate the weak magnetic scattering from the phonon scattering as outlined in Section IV.

### B. INS Experiments

In order to extend the range of energy transfers $\Delta E$ beyond that reported in Ref. [8], new INS experiments were carried out with use of the high-resolution direct geometry time-of-flight chopper spectrometer SEQUOIA [9] at the spallation neutron source (SNS) at Oak Ridge National Laboratory. The samples were enclosed in aluminum cylinders (8 mm diameter, 45 mm height) and placed into a closed-cycle He cryostat. Additional experiments were performed for vanadium to allow the correction of the raw data with respect to background, detector efficiency, and absorption according to standard procedures. The measurements were performed at T=5 K for incoming neutron energies $E_0$=80 and 120 meV, yielding instrumental energy resolutions of about 1.9% with respect to the energy of the scattered neutrons $E_1=E_0-\Delta E$.

### III. THEORETICAL BACKGROUND

#### A. The general $Cu^{2+}$ hexamer model for $A_2Cu_3O(SO_4)_3$ (A=Na, K)

The energies of the spin excitations observed for magnetic clusters are usually analyzed in terms of a Heisenberg spin Hamiltonian of the form

$$H = -2\sum_{i,j=1}^{6} J_{ij} \mathbf{S_i} \cdot \mathbf{S_j}, \tag{1}$$

where $J_{ij}$ and $\mathbf{S_i}$ denote the bilinear exchange parameters and the spin operators of the magnetic ions, respectively. For the compounds $A_2Cu_3O(SO_4)_3$ (A=Na, K), Eq. (1) gives rise to twenty $Cu^{2+}$ hexamer states, namely five singlets ($S_i$), nine triplets ($T_i$), five quintets ($Q_i$), and one septet. In principle, Eq. (1) involves fifteen independent exchange parameters $J_{ij}$, whose number can be reduced by several measures. The $Cu^{2+}$ hexamer has $C_2$ point symmetry, connecting the centers of the two $Cu_4$ tetrahedra, thus we have $J_{13}=J_{24}$, $J_{14}=J_{23}$, $J_{35}=J_{46}$, and $J_{36}=J_{45}$ (see Fig. 1). By confining to the leading nearest-neighbor interactions, we set $J_{15}=J_{16}=J_{25}=J_{26}=0$. Thus we end up with a total of seven independent exchange parameters $J_{12}$, $J_{13}$, $J_{14}$, $J_{34}$, $J_{35}$, $J_{36}$, and $J_{56}$. In hydroxo bridged $Cu^{2+}$ dimers it was found that with increasing bond angle the interaction switched from ferromagnetic to antiferromagnetic around 98° [10]. These empirical observations could be rationalized using simple orbital overlap considerations [11]. Based on the bonding data listed in Table I, we cannot expect to predict the coupling strength of a given $Cu^{2+}$ pair from the geometry. But with bond angles 91°<Θ<94° the $Cu_3$-O-$Cu_4$ bridge is standing out, thus we expect strong ferromagnetic exchange for $J_{34}$. For $J_{12}$ and $J_{56}$, with bond angles 102°<Θ<105° and 97°<Θ<102°, respectively, we expect a smaller interaction, either ferromagnetic or antiferromagnetic. For the remaining interactions it is plausible to expect the antiferromagnetic contribution to dominate.

## B. The four-parameter model for $A_2Cu_3O(SO_4)_3$ (A=Na, K)

The seven-parameter model results in a non-diagonal spin Hamiltonian (1), which is not beneficial for a deeper physical understanding of the $Cu^{2+}$ hexamers. In particular, the selection rules in INS experiments cannot be derived in a straightforward manner. We therefore make a simplification and assign the same value to all the antiferromagnetic exchange parameters: $J_{13}=J_{14}=J_{35}=J_{36}$, thus ending up with a model consisting of four parameters $J_{12}$, $J_{13}$, $J_{34}$, and $J_{56}$. By choosing the spin coupling scheme **$S_{12}=S_1+S_2$, $S_{56}=S_5+S_6$, $S_{1256}=S_{12}+S_{56}$, $S_{34}=S_3+S_4$**, and **$S_{tot}=S_{1256}+S_{34}$**, with $0 \leq (S_{12},S_{34},S_{56}) \leq 1$, $0 \leq S_{1256} \leq 2$, and $0 \leq S_{tot} \leq 3$, the spin Hamiltonian (1) becomes diagonal, so that we arrive at closed-form expressions for the eigenvalues:

$$E(S_{12},S_{56},S_{1256},S_{34},S_{tot}) = -J_{12}[S_{12}(S_{12}+1)-T] - J_{34}[S_{34}(S_{34}+1)-T]$$
$$- J_{56}[S_{56}(S_{56}+1)-T] + J_{13}[S_{1256}(S_{1256}+1)+S_{34}(S_{34}+1)-S_{tot}(S_{tot}+1)] , \qquad (2)$$

with $T=2S_i(S_i+1)$ and $S_i=1/2$ (i=1,2,...,6).

The eigenfunctions correspond to the basis functions $|S_{12},S_{56},S_{1256},S_{34},S_{tot}\rangle$. The results of the four-parameter model are summarized in Table II. Note that an unambiguous description of the spin states of an N-mer requires (N-2) additional spin quantum numbers besides $S_{tot}$.

### C. The five-parameter model for $A_2Cu_3O(SO_4)_3$ (A=Na, K)

It is meaningful to relax the constraint $J_{13}=J_{14}$ when considering the Cu-O-Cu bond angles and Cu-Cu distances listed in Table I. We define

$$J_{13} = J^0_{13,14}(1-\varepsilon) , \quad J_{14} = J^0_{13,14}(1+\varepsilon) , \qquad (3)$$

thus ending up with five parameters $J_{12}$, $J_{13}=J_{35}$, $J_{14}=J_{36}$, $J_{34}$, and $J_{56}$. This coupling scheme does no longer result in a diagonal energy matrix. The eigenvectors are linear combinations of the basis functions $|S_{12},S_{56},S_{1256},S_{34},S_{tot}\rangle$, so that the energy matrix of dimension $2^6=64$ must be solved numerically.

## D. Neutron cross-section for $A_2Cu_3O(SO_4)_3$ (A=Na, K)

The neutron cross-section for spin hexamer transitions can be derived from the general formula for magnetic neutron scattering [12,13]:

$$\frac{d^2\sigma}{d\Omega d\omega} \propto F^2(\mathbf{Q}) \sum_{\alpha\beta} \left( \delta_{\alpha\beta} - \frac{Q_\alpha Q_\beta}{Q^2} \right) S^{\alpha\beta}(\mathbf{Q},\omega)$$

with (4)

$$S^{\alpha\beta}(\mathbf{Q},\omega) = \sum_{i,j=1}^{6} \exp\{i\mathbf{Q} \cdot (\mathbf{R}_i - \mathbf{R}_j)\} \sum_{\lambda,\lambda'} p_\lambda \langle \lambda | S_i^\alpha | \lambda' \rangle \langle \lambda' | S_j^\beta | \lambda \rangle \, \delta(\hbar\omega + E_\lambda - E_{\lambda'})$$

where $F(\mathbf{Q})$ is the magnetic form factor, $\mathbf{Q}$ the scattering vector, and $S_i^\alpha$ ($\alpha$=x,y,z) the spin operator of the $i$th ion at site $\mathbf{R}_i$. $|\lambda\rangle$ denotes the initial state of the system, with energy $E_\lambda$ and thermal population factor $p_\lambda$, and its final state is $|\lambda'\rangle$. $\langle\lambda|S_i^\alpha|\lambda'\rangle$ and $\langle\lambda'|S_j^\beta|\lambda\rangle$ are the transition matrix elements. The eigenfunctions $|\lambda\rangle$ and $|\lambda'\rangle$ are expressed as linear combinations of the basis functions $|S_{12},S_{56},S_{1256},S_{34},S_{tot}\rangle$ described in Sections III.B and III.C.

For polycrystalline material Eq. (4) has to be averaged in $\mathbf{Q}$ space, which concerns the polarization factor ($\delta_{\alpha\beta}-Q_\alpha Q_\beta/Q^2$) and the structure factor $\exp\{i\mathbf{Q}\cdot(\mathbf{R}_i-\mathbf{R}_j)\}$. The Q-averaging procedure results in damped oscillatory Q-dependences of the intensities which for all transitions start at zero intensity for Q=0, have maxima in the range 0.7<Q<1.5 Å$^{-1}$, and decrease for Q>2 Å$^{-1}$ according to $F^2(Q)$ with some minor modulations.

## E. Selection Rules in INS Experiments for $A_2Cu_3O(SO_4)_3$ (A=Na, K)

The four-parameter model described in Section III.B readily reveals the selection rules in INS experiments. In the dipole approximation, the neutron can only produce ΔS=0 and ΔS=±1 transitions for each spin quantum number S, thus the primary selection rules are as follows:

$$\Delta S_{12}=0,\pm 1,\ \Delta S_{56}=0,\pm 1,\ \Delta S_{1256}=0,\pm 1,\ \Delta S_{34}=0,\pm 1,\ \Delta S_{tot}=0,\pm 1\ . \tag{5}$$

In addition, we have secondary selection rules due to the fact that the neutron cannot simultaneously excite more than one component of the hexamer, which constitutes further constraints to INS experiments [14]. According to the spin coupling scheme defined in Section III.B, the hexamer components are defined by three dimers ($Cu_1$-$Cu_2$, $Cu_3$-$Cu_4$, $Cu_5$-$Cu_6$) and a tetramer ($Cu_1$-$Cu_2$-$Cu_5$-$Cu_6$). Consequently the secondary selection rules are as follows:

$$|\Delta S_{12}|+|\Delta S_{56}|+|\Delta S_{34}|\leq 1,\ \ |\Delta S_{1256}|+|\Delta S_{34}|\leq 1\ . \tag{6}$$

With the triplet state $T_1$ ($S_{tot}=1$) being the ground state, only the transition to the septet state ($S_{tot}=3$) with $\Delta S_{tot}=2$ is not allowed according to the primary selection rules (Eq. 5). For the four-parameter model, however, the secondary selection (Eq. 6) rules forbid seven further transitions ($S_4$, $S_5$, and $T_5$-$T_9$) which are marked with zero intensity in the column $I_{calc}(\varepsilon=0)$ of Table II. In summary, the selection rules allow transitions to the singlet states $S_1$, $S_2$, $S_3$, to the triplet states $T_2$, $T_3$, $T_4$, and to all the quintet states $Q_1$-$Q_5$, the latter having very small intensities for $Q_2$, $Q_3$, and $Q_4$.

For the five-parameter model, the above mentioned selection rules are relaxed as shown in the column $I_{calc}(\varepsilon=0.3)$ of Table II. Transitions which are strictly forbidden in the four-parameter model, attain only very little intensity, so that we classify them as hidden transitions which cannot be detected in INS experiments on present-day neutron sources. This is basically different for the transitions $Q_1$-$Q_5$ which exhibit a substantial redistribution of the intensities for increasing values of ε.

## IV. EXPERIMENTAL RESULTS

The following presentation includes the data obtained with use of the high-resolution time-of-flight spectrometer CNCS at SNS Oak Ridge reported in Ref. [8] as well as the new experiments carried out with use of the spectrometer SEQUOIA (see Section II.B). Fig. 2 shows energy spectra observed for $A_2Cu_3O(SO_4)_3$, which exhibit similar features for both A=Na and A=K as expected from the similar structural parameters. The data correspond to the sum of magnetic and phonon scattering. Phonons here refer to vibrations of molecular solids, including both external (intermolecular) and internal (intramolecular) modes. The scattering is strongly dependent on the modulus of the scattering vector **Q** as exemplified in Fig. 3(a). With increasing Q, the intensity of the magnetic scattering decreases for Q>2 Å$^{-1}$ according to Eq. (4) with $F^2(Q)$ (apart from small modulations due to the structure factor), whereas the intensity of phonon scattering increases with $Q^2$ (apart from polarization factors entering the phonon cross-section [15]). We therefore conclude that the local maxima showing up around 15 and 30 meV (as well as the shoulder at 33 meV) correspond to magnetic excitations already identified in Ref. [8]. However, it is hard to identify further magnetic scattering for ΔE>35 meV.

By considering intensity ratios $I(Q_n)/I(Q_m)$ exemplified in Fig. 3(b), the magnetic scattering can readily be identified. The intensity ratio displayed at the top of Fig. 3(b) is almost constant over the whole energy range, starting from 0.8 at low energy transfers and smoothly increasing to about 0.9 at high energy transfers. This behavior is almost completely due to phonon scattering, since the magnetic form factor is very small for high Q values. On the other hand, the two intensity ratios at the bottom of Fig. 3(b) exhibit several peaks which can be attributed to magnetic scattering, since F(Q) is reasonably high at lower Q values. In particular, magnetic excitations are seen around $\Delta E=15$ meV ($S_1$, $S_2$, $S_3$ transitions) and at $25<\Delta E<40$ meV ($T_2$, $Q_1$, $T_3$, $T_4$ transitions). More importantly, there is evidence for magnetic transitions around $\Delta E=55$ and 65 meV, which can hardly be seen in Figs. 2 and 3(a).

A least-squares fitting procedure was developed to quantitatively derive both the magnetic and the phonon scattering from the energy spectra observed at different Q values for $Q>2$ Å$^{-1}$. The procedure is based on the specific Q-dependence of the magnetic and phonon scattering contributions:

$$S_{magnetic}(Q,\Delta E) = c \cdot F^2(Q) \cdot I_{magnetic}(\Delta E)$$

$$S_{phonon}(Q,\Delta E) = c \cdot R_{phonon}(Q, \Delta E) \cdot I_{phonon}(\Delta E) \tag{7}$$

The factor c contains several constants of the neutron cross-section as well as the Debye-Waller factor $e^{-2W(Q)}$. The factor $R_{phonon}(Q,\Delta E)$ is taken from the (smoothed) intensity ratios (see, *e.g.*, Fig. 3(b)) for $\Delta E$-ranges where no magnetic scattering is present, and it is interpolated for the magnetic $\Delta E$-ranges. The output functions are $I_{magnetic}(\Delta E)$ and $I_{phonon}(\Delta E)$, whose reliability strongly depends on the counting statistics. More specifically, the experimental error of the Q-dependent intensities should be of the order of a few percent, which was achieved for the data obtained by the instrument SEQUOIA. Consequently, the procedure works well for the data with $E_0=80$ and 120 meV. The data obtained by the instrument CNCS with $E_0=29.7$ and 50 meV, on the other hand, have Q-dependent intensities with an experimental error of about 10%, thus a different approach was adopted for the extraction of $I_{magnetic}(\Delta E)$ as outlined below.

The data obtained at $E_0=29.7$ meV for A=Na (see Fig. 2) exhibit three narrow lines of magnetic origin in the energy ranges $13<\Delta E<15$ meV ($S_1$, $S_2$) and 17<ΔE<19 meV ($S_3$) sitting on top of a large phonon background. The transition energies can readily be derived, but not their intensities. The data outside the above energy ranges due to phonon scattering were fitted with a polynomial function of 7$^{th}$ order as shown in Fig. 4(a). The difference between the observed data and the phonon scattering corresponds to the magnetic scattering as shown in Fig. 4(b). In the energy spectra observed at $E_0$=50 meV (see Fig. 2), the transitions $T_2$, $Q_1$, $T_3$, and $T_4$ cover a rather large energy range 24<ΔE<35 meV, so that we used as background the properly scaled phonon scattering taken from the data at $E_0$=80 meV as shown in Fig. 5. The same procedures were also applied for A=K.

The energy spectra obtained at $E_0$=80 and 120 meV were analyzed according to Eq. (7). This is exemplified for A=K ($E_0$=80 meV) and A=Na ($E_0$=120 meV) in Figs. 6 and 7, respectively. The transitions $Q_2$-$Q_4$ appear in a saddle of the phonon spectrum around ΔE=55 meV, whereas the transition $Q_5$ coincides with a local maximum of the phonon spectrum. This nicely demonstrates that our procedure based on intensity ratios is indispensable to reliably extract small magnetic signals from energy spectra dominated by phonon scattering.

Table III provides a summary of the transition energies observed at T=5 K for both A=Na and A=K. Note that part of our INS experiments involving the transitions $T_2$, $Q_1$, $T_3$, and $T_4$ were performed at T=1.5 K, *i.e.*, below the onset of magnetic order at $T_N$≈3 K. As shown in Ref. [8], the transition energies experience downward shifts of 0.2 meV upon lowering the temperature from T=5 K to T=1.5 K, so that the transition energies listed in Table III are corrected accordingly.

## V. DATA ANALYSIS

Least-squares fitting procedures were applied to the excitation energies observed for $A_2Cu_3O(SO_4)_3$ (A=Na, K) as listed in Table III. The resulting exchange parameters and the calculated energy spectra are summarized in Table IV and in Fig. 8, respectively. The fits based on the four- and five-parameter models (Sections III.B and III.C) gave a rather good agreement between the observed and calculated energies. However, the intensities associated with the transitions $Q_2$-$Q_5$ could not be satisfactorily reproduced as shown in Fig. 8. This is no surprise for the four-parameter model, which predicts extremely small intensities for the transitions $Q_2$-$Q_4$ listed in Table II by the column $I_c(\varepsilon=0)$. For the five-parameter model ($0.13 \leq \varepsilon \leq 0.16$), the intensity discrepancy between the transitions $Q_2$-$Q_4$ and $Q_5$ is only marginally remedied. What is needed are much larger values of $\varepsilon$ as shown in Table II by the column $I_c(\varepsilon=0.3)$. Therefore, the final fits were carried out for the seven-parameter model (Section III.A) which indeed provided $\varepsilon$-values close to 0.3 with reasonable intensities for the transitions $Q_2$-$Q_5$. For the intensity calculations we used the program MAGPACK developed by Borras-Almenar *et al.* [16].

The analysis of the Q-dependence of the intensities is a powerful tool for the correct assignment of the transitions. This is demonstrated in Fig. 9 for the $S_1$, $S_2$, and $S_3$ transitions observed for $A_2Cu_3O(SO_4)_3$ (A=Na, K). The cross-section calculated for the $S_1$ transition is very different from that calculated for the $S_2$ and $S_3$ transitions, which is nicely confirmed by the experimental data.

## VI. DISCUSSION AND CONCLUDING REMARKS

We presented INS data for the compounds $A_2Cu_3O(SO_4)_3$ (A=Na, K) and analyzed them in terms of spin-coupling parameters. They confirm the triplet S=1 ground state and provide an excellent agreement with the magnetic susceptibility data up to T=300 K [7] by using a g-factor g=2.05. The energy range of the magnetic excitations extends up to 70 meV (the corresponding temperature scale is about 800 K), which contrasts to the antiferromagnetic transition temperature $T_N$=3 K. This means that the magnetic properties at low temperatures are essentially determined by the ground-state triplet alone whose wave-functions, however, decisively depend on the excitation energies of all the Cu hexamer states.

An outstanding feature of the present study is the selective distribution of the total INS intensity on the transitions. In both title compounds the $Cu_6$ cluster has $C_2$ point symmetry, the twofold axis being the long cluster axis. There are no symmetry based selection rules for INS transitions. As a result, there are a total of eighteen allowed transitions out of the triplet ground-state $T_1$. Experimentally, however, only nine bands are observed: In the low energy region (E<40 meV), six bands are assigned to each the spin singlet ($S_1$, $S_2$, $S_3$) and triplet ($T_2$, $T_3$, $T_4$) excitations, and one to the quintet $Q_1$ excitation. The two high-energy bands around 55 meV and 65 meV are composed of partially unresolved $Q_2$, $Q_3$, $Q_4$, and $Q_5$ excitations, respectively. Inspection of Table II reveals that, using the five-parameter model, the seven unobserved allowed transitions $S_4$, $S_5$, and $T_5$-$T_9$ have calculated intensities which are typically two orders of magnitude smaller than for the observed ones. This is significant and worth exploring. There is a mechanism at work which distributes the total intensity very unequally on the allowed transitions.

We believe that the origin of this phenomenon lies in the topology of the spin coupling in this cluster. We refer to the four-parameter model (Section III.B), in which we have artificially introduced a mirror plane, thus raising the point symmetry from the actual $C_2$ to $C_{2v}$. In this model the interactions both within the central $Cu^{2+}$ pair ($J_{34}$) and the two terminal $Cu^{2+}$ pairs ($J_{12}$, $J_{56}$) are ferromagnetic, whereas the interactions between the two central $Cu^{2+}$ ions and the four terminal $Cu^{2+}$ ions are all antiferromagnetic and equal (see Table IV). The latter is a drastic simplification, although setting $J_{13}=J_{24}=J_{35}=J_{46}$ can be justified by symmetry and the very similar Cu-O-Cu bridging geometries of the four connections, see Table I. The same is true for $J_{14}=J_{23}=J_{45}=J_{36}$. But assigning the same J value to both groups is a gross approximation. In the former group the Cu-O-Cu bond angles are much smaller than in the latter group (see Table I), thus we expect stronger antiferromagnetic exchange for the latter group.

The intensity distribution of the observed transitions is surprisingly well reproduced by the four-parameter model as demonstrated in Figs. 4-8, with the notable exception of the high-energy bands around 55 and 65 meV. How is this possible, with the drastic approximation of setting all eight antiferromagnetic exchange parameters equal? The cluster wavefunctions in the four-parameter model must be good approximations of the proper cluster wavefunctions. We ascribe this to the fact that in going from the five-parameter to the four-parameter model the coupling topology remains intact: parameter values change, the antiferromagnetic ones quite considerably, but they preserve their sign.

The magnetic nature of the compounds $A_2Cu_3O(SO_4)_3$ (A=Na, K) below $T_N$ is not yet fully settled. From the structural point-of-view there is no doubt that the copper hexamers form distinct layers parallel to the yz-plane [4,5,17,18]. Planes containing A ions are located between these layers, which manifests itself in a 10% difference of the lattice constants a(K)>a(Na). Furthermore, the good cleavage [100] of single crystals is fully explained by the layered structure [17]. Along both the y- and z-directions, the copper hexamers are connected through Cu-O-S-O-Cu bridges with typical bond lengths $d(Cu_i-Cu_j) \approx 4.3$ Å, whereas $d(Cu_i-Cu_j) \approx 7$ Å along the x-direction. Therefore it is reasonable to assume that the copper hexamers form a weakly, antiferromagnetically coupled two-dimensional magnetic network, which contrasts to the picture of one-dimensional magnetic chains postulated in Ref. [7]. A λ-type peak was observed in the temperature derivative of the magnetic susceptibility at $T_N$=3.1 K [6] typical of a second-order phase transition. The observation of magnetic Bragg peaks in neutron diffraction experiments performed below $T_N$ indicate the presence of long-range magnetic order with propagation vector **k**=(0,0,0) [6]. However, the observed magnetic reflections were rather weak, thus further neutron diffraction measurements with improved statistics and preferably on single crystals are highly desirable.

In conclusion, we have shown that extending the experimental energy range of the INS experiments beyond that reported in Refs. [7,8] is essential for a proper understanding of the exchange couplings in the title compounds. The data analysis turned out to be complicated by the strong phonon scattering superimposed on the rather weak magnetic scattering in the high-energy range. By developing a procedure based on the distinctly different Q-dependence of the two scattering contributions, however, it was possible to extract the magnetic signals with reasonable statistics.

An important part of the present work was devoted to an in-depth consideration of INS intensities, which had been neglected before. The very selective distribution of INS intensities was a key to understand the importance of the exchange topology in this system. The hidden selection rules can be correlated with true selection rules in a model, which is greatly simplified but retains the coupling topology. Least-squares fitting of both INS peak energies and intensities required the development of new software. It turned out that the observed experimental intensities could best be reproduced using such a procedure.

**ACKNOWLEDGMENTS**

We thank V. Pomjakushin (Paul Scherrer Institut, Switzerland) for useful discussions on the structural aspects. J.M.C. acknowledges the support by the Spanish MINECO (CTQ2017-89528-P). This research used resources at the

TABLE I. $Cu_i$-O-$Cu_j$ bond angles ($\Theta$) and $Cu_i$-$Cu_j$ bond distances (d) of the compounds $A_2Cu_3O(SO_4)_3$ (A=Na, K) determined at T=2 K by neutron diffraction [8].

| i-j | $\Theta(Cu_i$-O-$Cu_j)$ [°] | $d(Cu_i$-$Cu_j)$ [Å] | $\Theta(Cu_i$-O-$Cu_j)$ [°] | $d(Cu_i$-$Cu_j)$ [Å] |
|---|---|---|---|---|
| | $Na_2Cu_3O(SO_4)_3$ | | $K_2Cu_3O(SO_4)_3$ | |
| 1-2 | 105.2(4) | 2.984(6) | 102.2(5) | 3.000(6) |
| 1-3, 2-4 | 106.5(4) | 3.084(6) | 107.0(5) | 3.118(8) |
| 1-4, 2-3 | 124.4(4) | 3.403(6) | 124.2(5) | 3.428(7) |
| 3-4 | 92.7(4) / 91.2(4) | 2.815(7) | 93.0(5) / 94.0(5) | 2.854(7) |
| 3-5, 4-6 | 110.6(4) | 3.179(6) | 111.7(5) | 3.226(8) |
| 3-6, 4-5 | 121.3(5) | 3.371(7) | 123.0(6) | 3.427(8) |
| 5-6 | 101.7(4) | 2.982(7) | 96.8(5) | 2.890(7) |

Table II: Spin quantum numbers $S_{12}$, $S_{56}$, $S_{1256}$, $S_{34}$, $S_{tot}$ and energies $E(J_{12},J_{13},J_{34},J_{56})$ for the four-parameter model (see Section III.B). The state $T_1$ is taken as the ground state. The energies $E_c(\varepsilon)$ and intensities $I_c(\varepsilon)$ were calculated for the four-parameter ($\varepsilon=0$) and five-parameter ($\varepsilon=0.3$) models with $J_{12}=1.2$, $J_{13}=-6.7$, $J_{34}=12.5$, and $J_{56}=2.3$ meV. The intensities $I_c(\varepsilon)$ were integrated for $1.3 \leq Q \leq 7.0$ Å$^{-1}$. Eq. (3) applies for the calculations with $\varepsilon=0.3$.

| State | $S_{12}$ | $S_{56}$ | $S_{1256}$ | $S_{34}$ | $S_{tot}$ | $E(J_{12},J_{13},J_{34},J_{56})$ | $E_c(\varepsilon=0)$ [meV] | $I_c(\varepsilon=0)$ | $E_c(\varepsilon=0.3)$ | $I_c(\varepsilon=0.3)$ |
|---|---|---|---|---|---|---|---|---|---|---|
| $S_1$ | 1 | 1 | 1 | 1 | 0 | $-2J_{13}$ | 13.60 | 5.443 | 14.03 | 5.298 |
| $S_2$ | 0 | 1 | 1 | 1 | 0 | $2J_{12}-2J_{13}$ | 14.80 | 5.986 | 14.42 | 6.164 |
| $S_3$ | 1 | 0 | 1 | 1 | 0 | $-2J_{13}+2J_{56}$ | 18.00 | 5.986 | 17.99 | 5.938 |
| $S_4$ | 1 | 1 | 0 | 0 | 0 | $-6J_{13}+2J_{34}$ | 65.00 | 0 | 66.31 | 0.004 |
| $S_5$ | 0 | 0 | 0 | 0 | 0 | $2J_{12}-6J_{13}+2J_{34}+2J_{56}$ | 70.60 | 0 | 73.35 | 0.006 |
| $T_1$ | 1 | 1 | 2 | 1 | 1 | 0 | 0 | 29.485 | 0 | 29.158 |
| $T_2$ | 1 | 1 | 1 | 1 | 1 | $-4J_{13}$ | 27.20 | 4.082 | 27.22 | 4.267 |
| $T_3$ | 0 | 1 | 1 | 1 | 1 | $2J_{12}-4J_{13}$ | 28.40 | 4.489 | 28.71 | 4.571 |
| $T_4$ | 1 | 0 | 1 | 1 | 1 | $-4J_{13}+2J_{56}$ | 31.60 | 4.489 | 32.00 | 4.398 |
| $T_5$ | 1 | 1 | 0 | 1 | 1 | $-6J_{13}$ | 40.80 | 0 | 41.14 | 0.014 |
| $T_6$ | 0 | 0 | 0 | 1 | 1 | $2J_{12}-6J_{13}+2J_{56}$ | 46.40 | 0 | 45.96 | 0.025 |
| $T_7$ | 1 | 1 | 1 | 0 | 1 | $-6J_{13}+2J_{34}$ | 65.00 | 0 | 66.75 | 0.000 |
| $T_8$ | 0 | 1 | 1 | 0 | 1 | $2J_{12}-6J_{13}+2J_{34}$ | 66.20 | 0 | 68.55 | 0.016 |
| $T_9$ | 1 | 0 | 1 | 0 | 1 | $-6J_{13}+2J_{34}+2J_{56}$ | 69.40 | 0 | 72.06 | 0.055 |
| $Q_1$ | 1 | 1 | 2 | 1 | 2 | $-4J_{13}$ | 27.20 | 24.105 | 28.02 | 23.826 |
| $Q_2$ | 1 | 1 | 1 | 1 | 2 | $-8J_{13}$ | 54.40 | 0.272 | 54.40 | 0.229 |
| $Q_3$ | 0 | 1 | 1 | 1 | 2 | $2J_{12}-8J_{13}$ | 55.60 | 0.299 | 55.22 | 3.717 |
| $Q_4$ | 1 | 0 | 1 | 1 | 2 | $-8J_{13}+2J_{56}$ | 58.80 | 0.299 | 58.53 | 0.734 |
| $Q_5$ | 1 | 1 | 2 | 0 | 2 | $-6J_{13}+2J_{34}$ | 65.00 | 10.843 | 68.94 | 7.262 |
| Septet | 1 | 1 | 2 | 1 | 3 | $-10J_{13}$ | 68.00 | 0 | 68.82 | 0 |

Table III: Excitation energies observed for $A_2Cu_3O(SO_4)_3$ (A=Na,K) at T=5 K.

| State | $Na_2Cu_3O(SO_4)_3$ $E_{obs}$ [meV] | $K_2Cu_3O(SO_4)_3$ $E_{obs}$ [meV] |
|---|---|---|
| $S_1$ | 13.5±0.2 | 12.6±0.2 |
| $S_2$ | 14.7±0.2 | 15.1±0.2 |
| $S_3$ | 18.0±0.2 | 18.0±0.3 |
| $T_2$ | 25.8±0.5 | 24.7±0.4 |
| $Q_1$ | 27.2±0.2 | 26.2±0.2 |
| $T_3$ | 29.3±0.4 | 27.8±0.4 |
| $T_4$ | 32.3±0.4 | 31.2±0.3 |
| $Q_2$-$Q_4$ | 55.7±0.5 | 54.9±0.5 |
| $Q_5$ | 65.3±2.0 | 65.7±2.0 |

Table IV: Exchange parameters $J_{ij}$ [meV] obtained for $A_2Cu_3O(SO_4)_3$ (A=Na,K). The exchange parameters $J_{ij}(4)$, $J_{ij}(5)$, and $J_{ij}(7)$ result from fitting the observed energies (listed in Table III) in the four-, five-, and seven-parameter models, respectively.

| $J_{ij}$ | $Na_2Cu_3O(SO_4)_3$ | | | $K_2Cu_3O(SO_4)_3$ | | |
|---|---|---|---|---|---|---|
| | $J_{ij}(4)$ | $J_{ij}(5)$ | $J_{ij}(7)$ | $J_{ij}(4)$ | $J_{ij}(5)$ | $J_{ij}(7)$ |
| $J_{12}$ | 0.6 ± 0.2 | 0.8 ± 0.1 | 1.3 ± 0.2 | 1.1 ± 0.2 | 1.2 ± 0.2 | 1.6 ± 0.4 |
| $J_{13}=J_{24}$ | -6.8 ± 0.1 | -5.7 ± 0.2 | -4.7 ± 0.3 | -6.5 ± 0.1 | -5.6 ± 0.2 | -4.6 ± 0.3 |
| $J_{14}=J_{23}$ | -6.8 ± 0.1 | -7.8 ± 0.2 | -8.3 ± 0.3 | -6.5 ± 0.1 | -7.3 ± 0.2 | -7.9 ± 0.3 |
| $J_{34}$ | 12.1 ± 1.5 | 12.0 ± 1.5 | 11.5 ± 1.5 | 13.4 ± 2.2 | 13.4 ± 1.4 | 12.4 ± 1.5 |
| $J_{35}=J_{46}$ | -6.8 ± 0.1 | -5.7 ± 0.2 | -5.3 ± 0.3 | -6.5 ± 0.1 | -5.6 ± 0.2 | -5.2 ± 0.3 |
| $J_{36}=J_{45}$ | -6.8 ± 0.1 | -7.8 ± 0.2 | -8.3 ± 0.3 | -6.5 ± 0.1 | -7.3 ± 0.2 | -7.9 ± 0.3 |
| $J_{56}$ | 2.2 ± 0.2 | 2.4 ± 0.2 | 2.2 ± 0.2 | 2.6 ± 0.2 | 2.6 ± 0.2 | 2.5 ± 0.3 |

**FIGURE CAPTIONS**

FIG. 1. (Color online) Schematic structure of the $Cu^{2+}$ hexamers in the compounds $A_2Cu_3O(SO_4)_3$ (A=Na, K). The independent ferromagnetic and antiferromagnetic exchange parameters $J_{ij}$ are marked with full and broken arrows, respectively, with the symmetry constraints listed at the bottom.

FIG. 2. (Color online) Energy spectra of neutrons scattered from $A_2Cu_3O(SO_4)_3$ for (a) A=Na and (b) A=K. The data were collected over the whole angular range provided by the CNCS and SEQUOIA spectrometers. For clarity, the data for $E_0$=29.7, 50, and 80 meV are enhanced by some intensity units. T=1.5 K for $E_0$=29.7 and 50 meV (CNCS). T=5 K for $E_0$=80 and 120 meV (SEQUOIA). $S_i$, $T_i$, and $Q_1$ mark the transitions of magnetic origin.

FIG. 3. (Color online) (a) Q-dependence of the energy spectra observed for $Na_2Cu_3O(SO_4)_3$ at T=5 K with $E_0$=120 meV. (b) Intensity ratios calculated from the data displayed in (a), denoted by $I(Q_n)/I(Q_m)$ in the text.

FIG. 4. (Color online) (a) Analysis of the data observed for $Na_2Cu_3O(SO_4)_3$ for $E_0$=29.7 meV. The dashed line corresponds to phonon scattering obtained by fitting the data outside the magnetic transitions with a polynomial function of $7^{th}$ order. (b) Extracted magnetic scattering as explained in the text. The lines are the results of a Gaussian least-squares fit without any constraints concerning energy transfers, intensities, linewidths, and background.

FIG. 5. (Color online) (a) Analysis of the data observed for $Na_2Cu_3O(SO_4)_3$ for $E_0$=50 meV. The square symbols correspond to phonon scattering as explained in the text. (b) Extracted magnetic scattering. The lines are as in Fig. 4(b).

FIG. 6. (Color online) Decomposition of the data observed for $K_2Cu_3O(SO_4)_3$ ($E_0$=80 meV) into magnetic and phonon scattering contributions. The full line is the result of a Gaussian least-squares fit.

FIG. 7. (Color online) Decomposition of the data observed for $Na_2Cu_3O(SO_4)_3$ ($E_0$=120 meV) into magnetic and phonon scattering contributions. The full and dashed lines are as in Fig. 4(b).

FIG. 8. (Color online) Energy spectra of $A_2Cu_3O(SO_4)_3$ calculated for A=Na (upper panel) and A=K (lower panel). $I_{calc}(4)$, $I_{calc}(5)$, and $I_{calc}(7)$ refer to energy fits obtained for the four-, five-, and seven-parameter model, respectively. The vertical bars mark the intensity amplitudes of the transitions (for A=Na taken from Figs. 4, 5, and 7). The integrated intensity of a transition is obtained by multiplying the amplitude with the linewidth. The error bars of the intensity amplitudes amount to about 30%.

FIG. 9. (Color online) Q-dependence of the intensities of the transitions $S_1$, $S_2$, and $S_3$ observed for $A_2Cu_3O(SO_4)_3$ ($E_0$=29.7 meV, T=1.5 K) with A=Na (circles) and A=K (squares). The lines correspond to the intensities calculated for the five-parameter model.

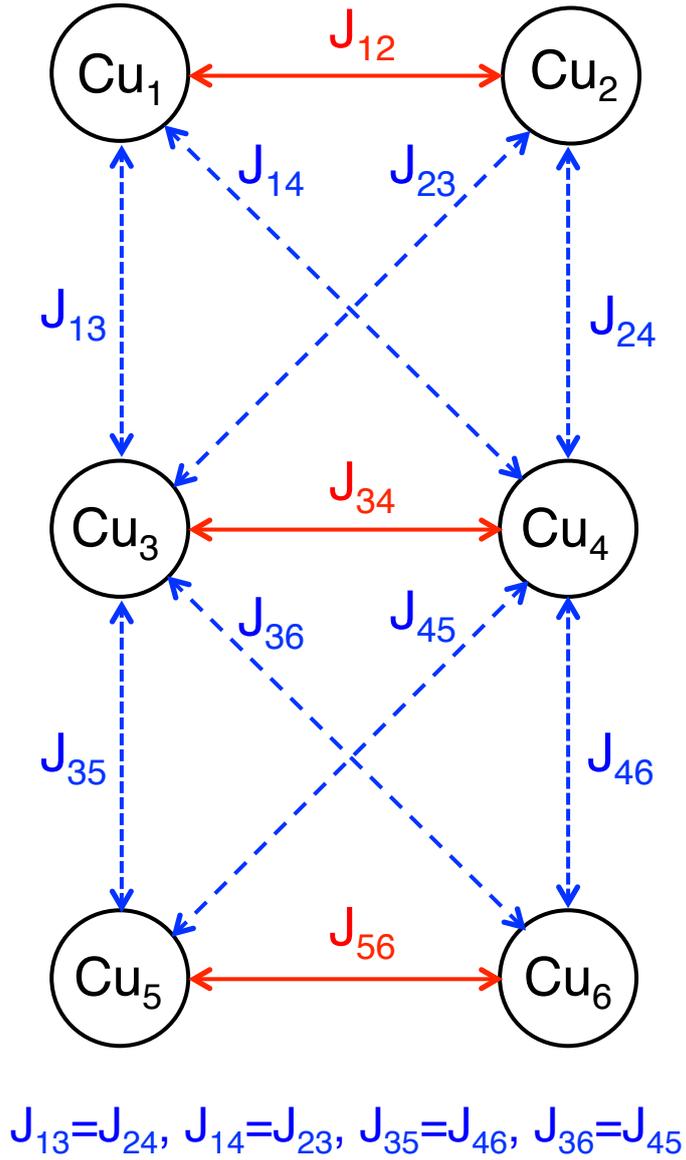

FIG. 1. (Color online) Schematic structure of the $Cu^{2+}$ hexamers in the compounds $A_2Cu_3O(SO_4)_3$ (A=Na, K). The independent ferromagnetic and antiferromagnetic exchange parameters $J_{ij}$ are marked with full and broken arrows, respectively, with the symmetry constraints listed at the bottom.

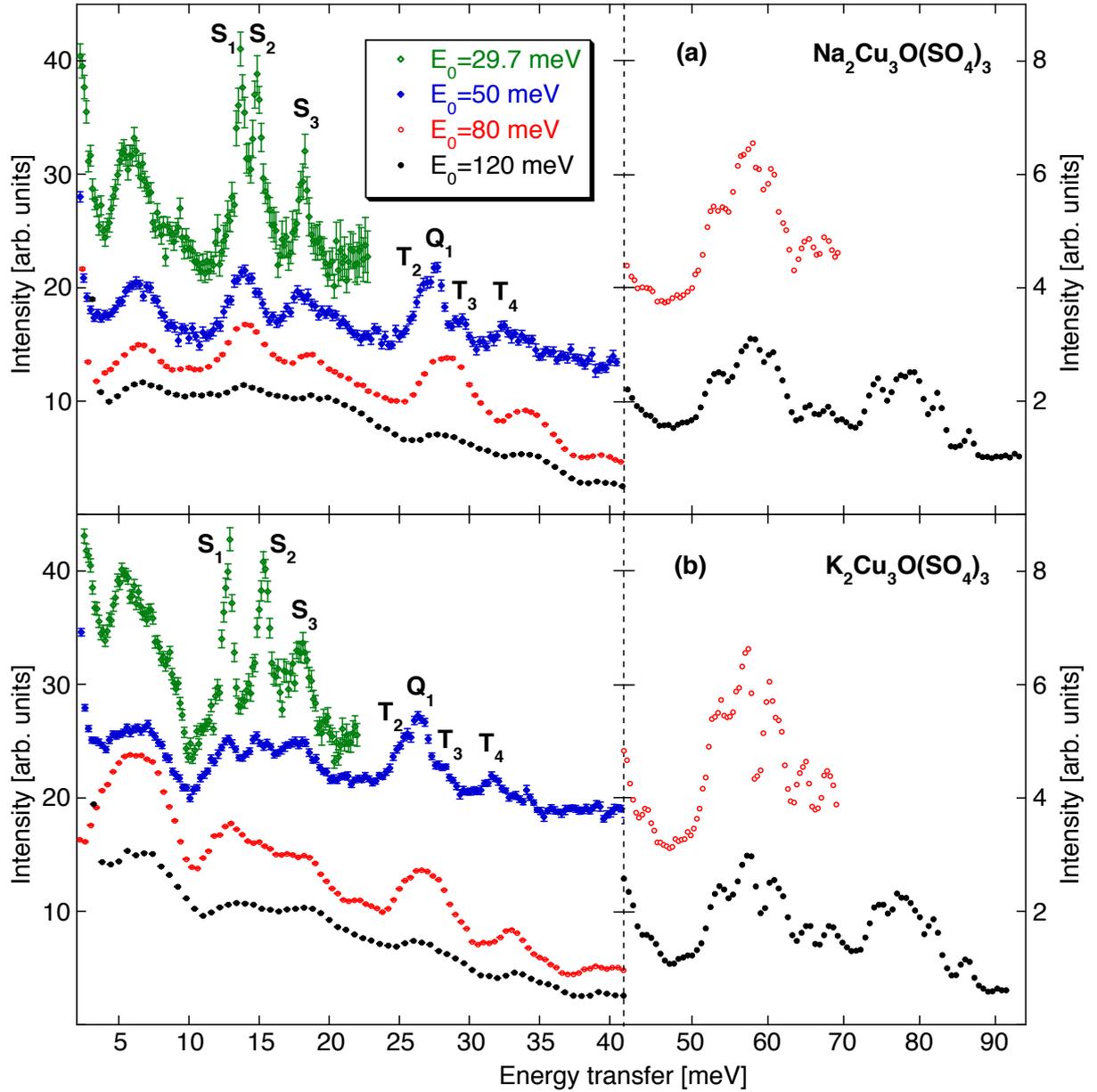

FIG. 2. (Color online) Energy spectra of neutrons scattered from $A_2Cu_3O(SO_4)_3$ for (a) A=Na and (b) A=K. The data were collected over the whole angular range provided by the CNCS and SEQUOIA spectrometers. For clarity, the data for $E_0$=29.7, 50, and 80 meV are enhanced by some intensity units. T=1.5 K for $E_0$=29.7 and 50 meV (CNCS). T=5 K for $E_0$=80 and 120 meV (SEQUOIA). $S_i$, $T_i$, and $Q_1$ mark the transitions of magnetic origin.

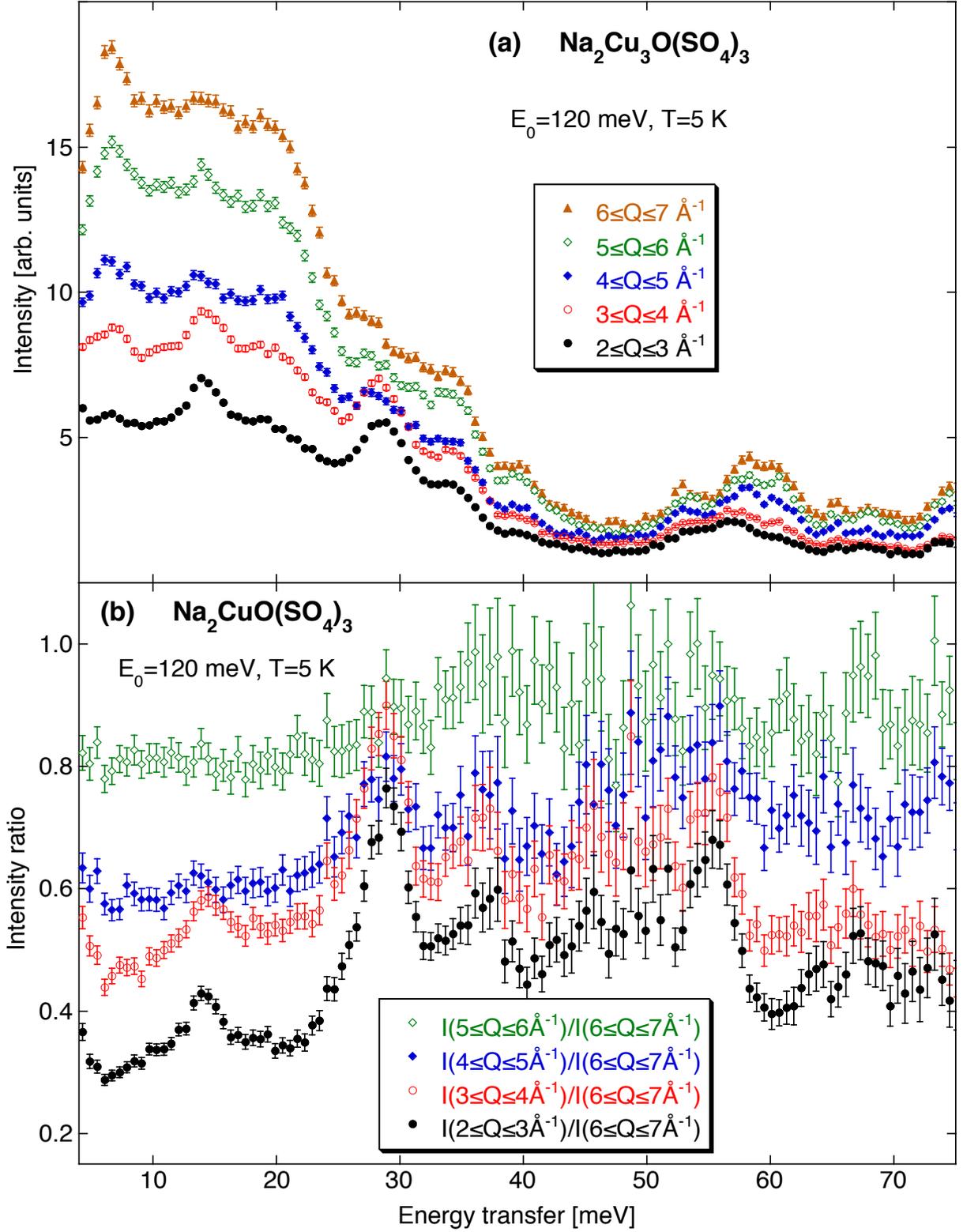

FIG. 3. (Color online) (a) Q-dependence of the energy spectra observed for $Na_2Cu_3O(SO_4)_3$ at T=5 K with $E_0$=120 meV. (b) Intensity ratios calculated from the data displayed in (a), denoted by $I(Q_n)/I(Q_m)$ in the text.

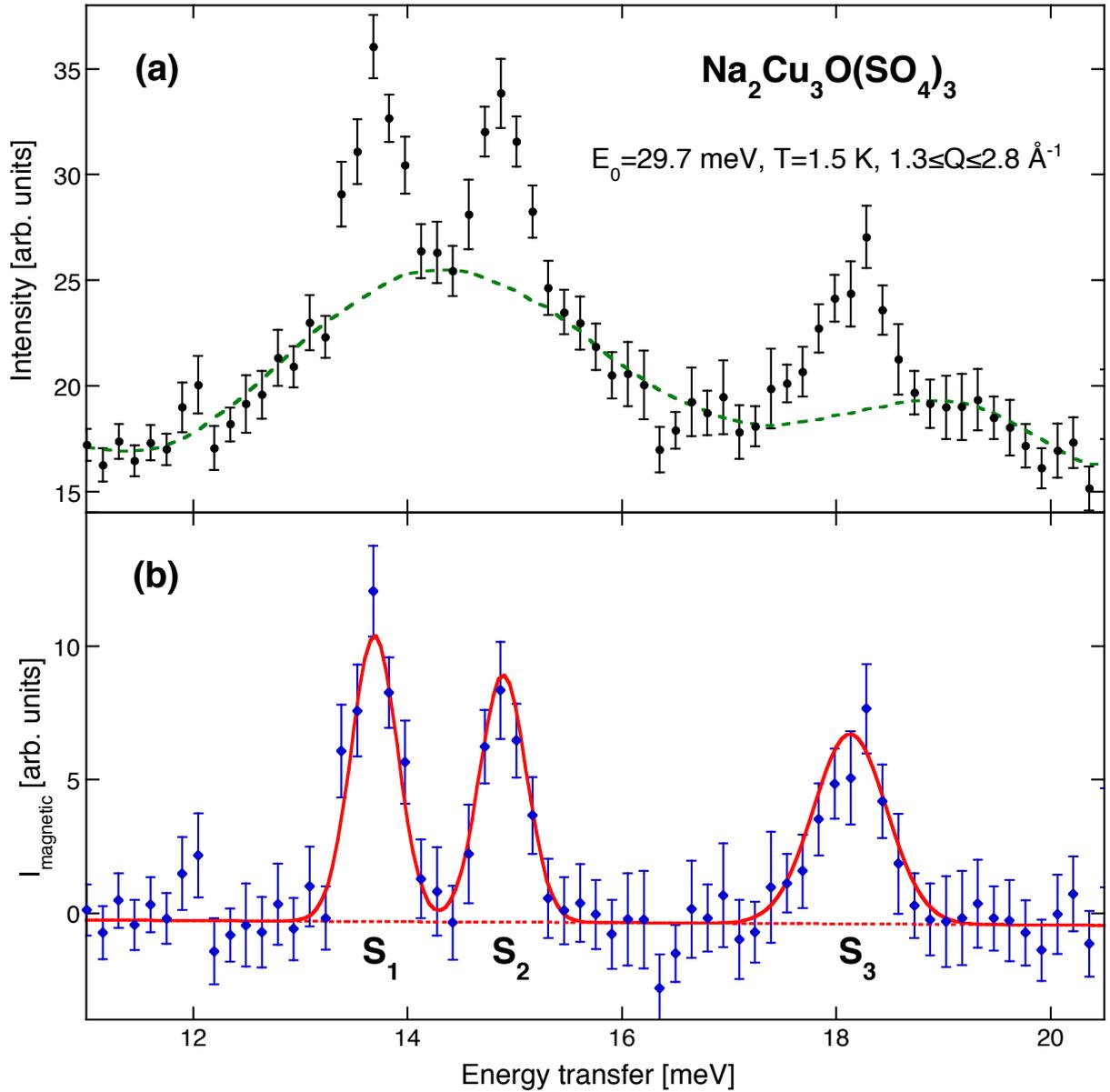

FIG. 4. (Color online) (a) Analysis of the data observed for $Na_2Cu_3O(SO_4)_3$ for $E_0$=29.7 meV. The dashed line corresponds to phonon scattering obtained by fitting the data outside the magnetic transitions with a polynomial function of $7^{th}$ order. (b) Extracted magnetic scattering as explained in the text. The lines are the results of a Gaussian least-squares fit without any constraints concerning energy transfers, intensities, linewidths, and background.

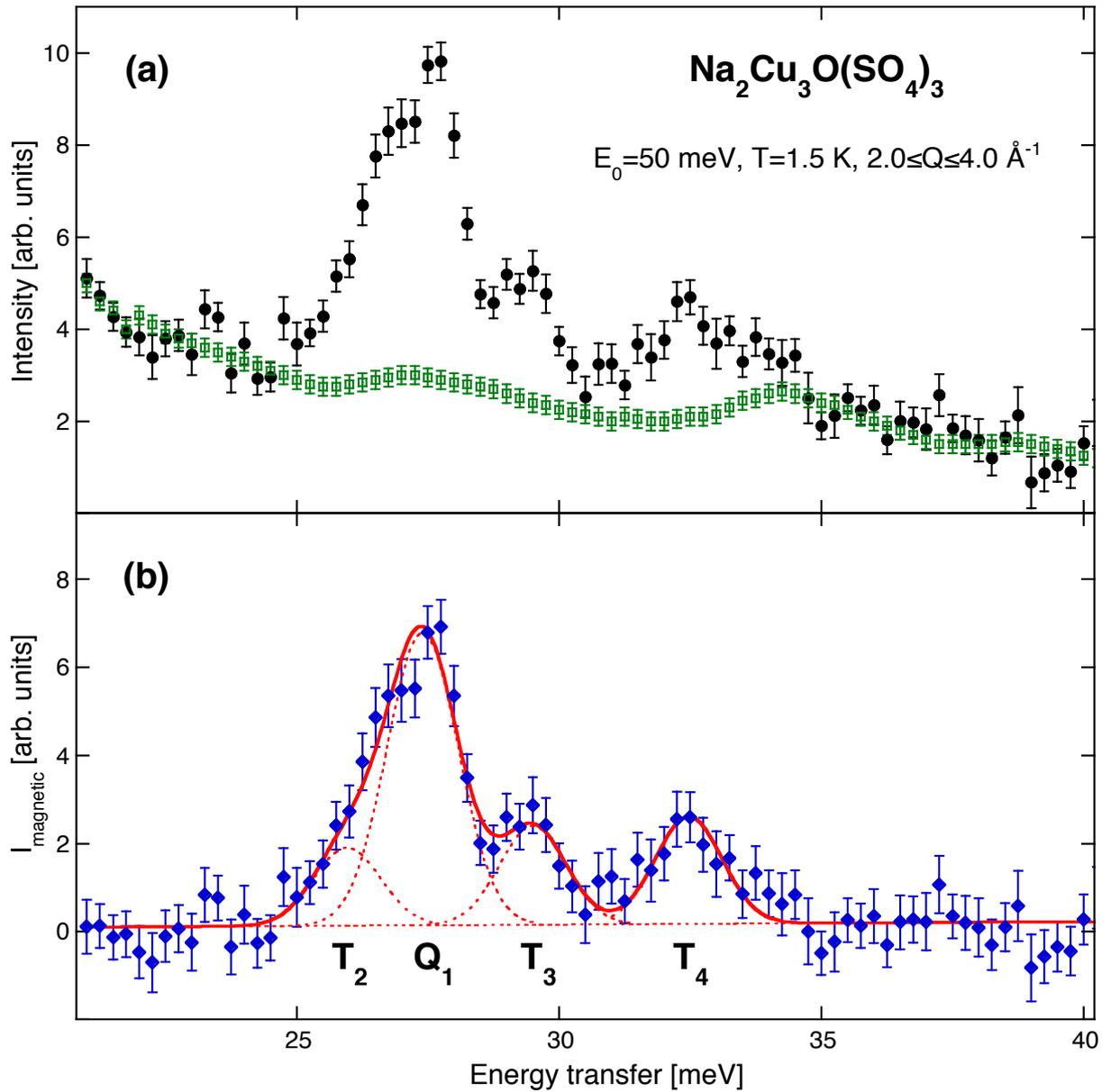

FIG. 5. (Color online) (a) Analysis of the data observed for Na$_2$Cu$_3$O(SO$_4$)$_3$ for E$_0$=50 meV. The square symbols correspond to phonon scattering as explained in the text. (b) Extracted magnetic scattering. The lines are as in Fig. 4(b).

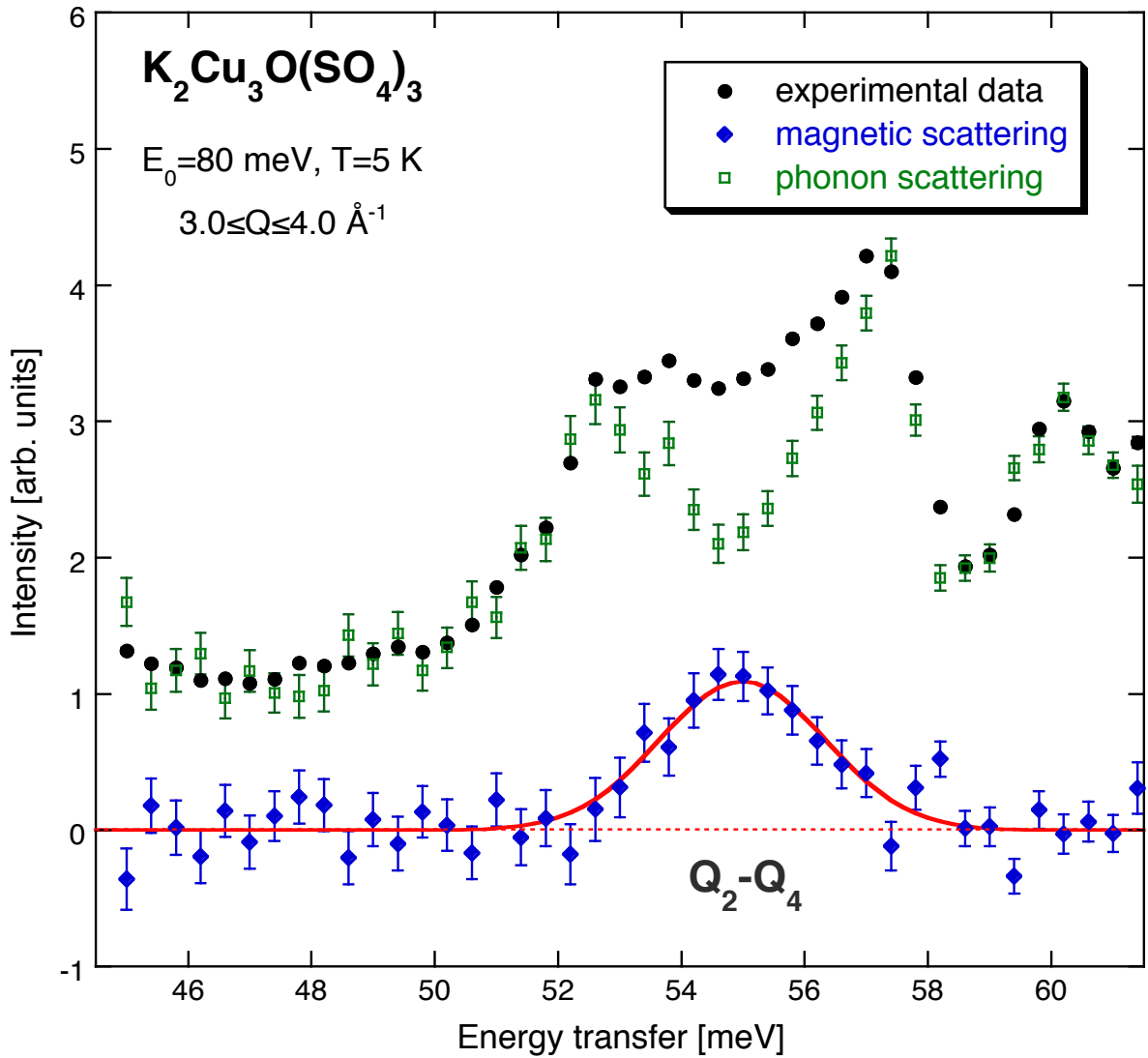

FIG. 6. (Color online) Decomposition of the data observed for $K_2Cu_3O(SO_4)_3$ ($E_0$=80 meV) into magnetic and phonon scattering contributions. The full line is the result of a Gaussian least-squares fit.

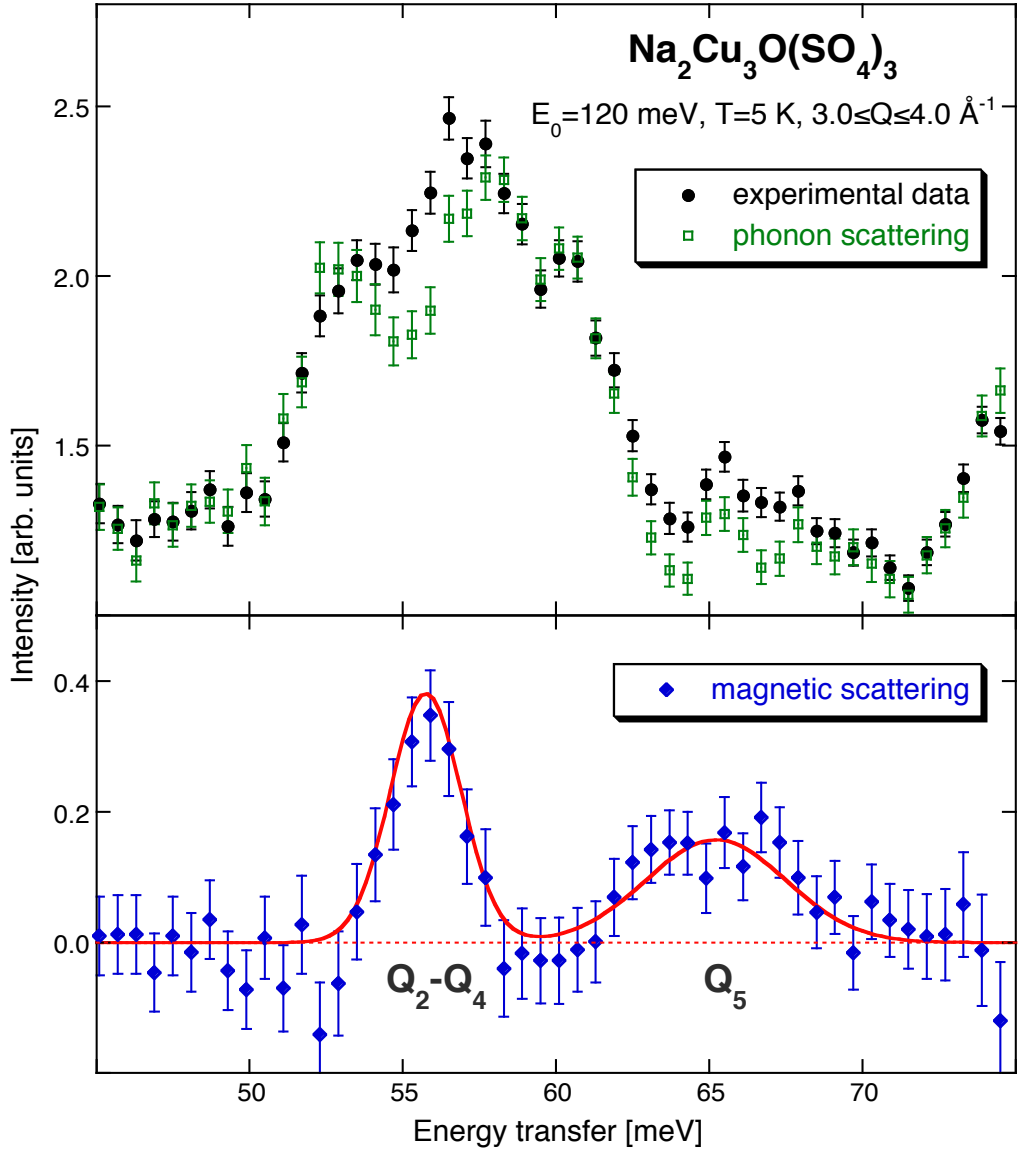

FIG. 7. (Color online) Decomposition of the data observed for $Na_2Cu_3O(SO_4)_3$ ($E_0$=120 meV) into magnetic and phonon scattering contributions. The full and dashed lines are as in Fig. 4(b).

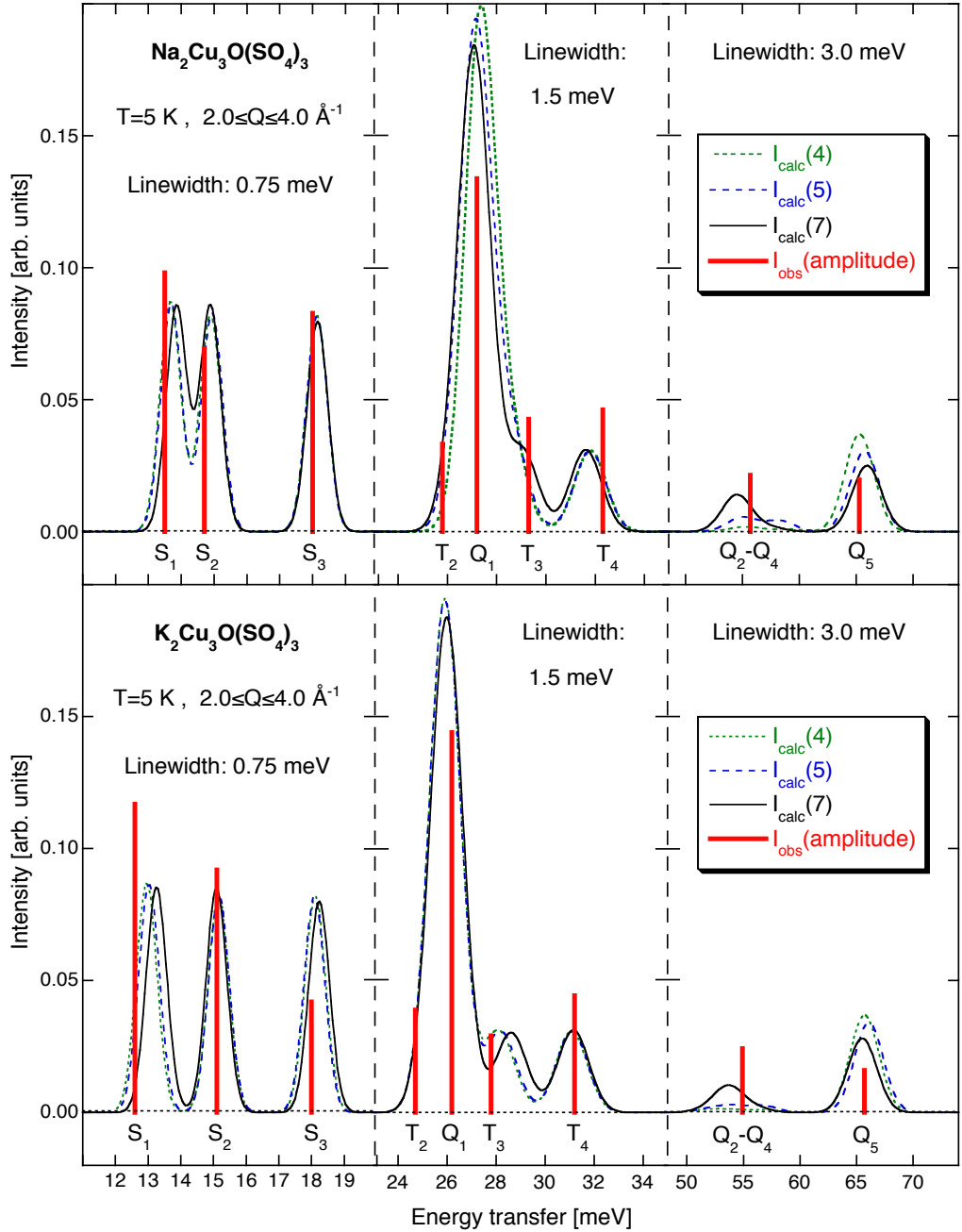

FIG. 8. (Color online) Energy spectra of $A_2Cu_3O(SO_4)_3$ calculated for A=Na (upper panel) and A=K (lower panel). $I_{calc}(4)$, $I_{calc}(5)$, and $I_{calc}(7)$ refer to energy fits obtained for the four-, five-, and seven-parameter model, respectively. The vertical bars mark the intensity amplitudes of the transitions (for A=Na taken from Figs. 4, 5, and 7). The integrated intensity of a transition is obtained by multiplying the amplitude with the linewidth. The error bars of the intensity amplitudes amount to about 30%.

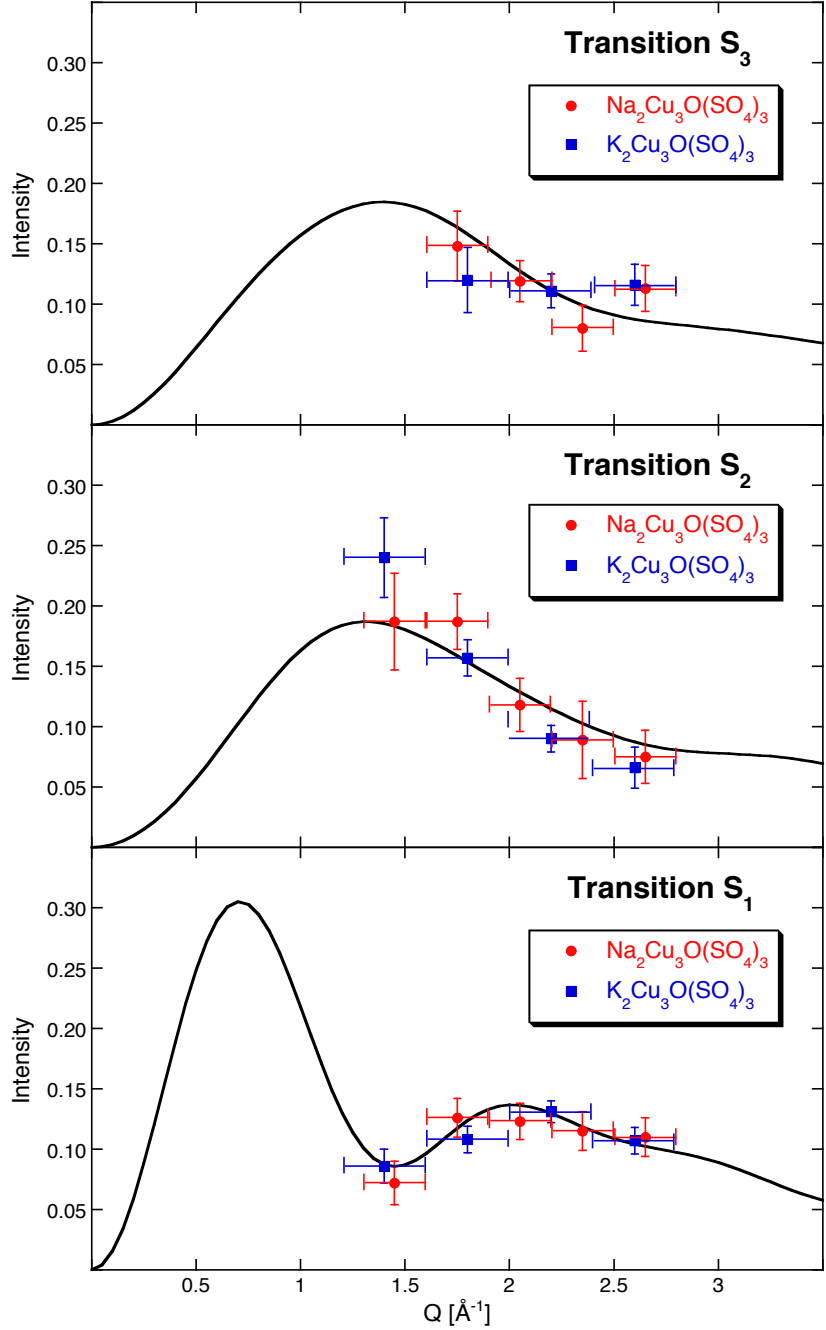

FIG. 9. (Color online) Q-dependence of the intensities of the transitions $S_1$, $S_2$, and $S_3$ observed for $A_2Cu_3O(SO_4)_3$ ($E_0$=29.7 meV, T=1.5 K) with A=Na (circles) and A=K (squares). The lines correspond to the intensities calculated for the five-parameter model.